\input{aipcheck}

\documentclass[
    ,final            
  ]
  {aipproc}

\layoutstyle{6x9}

\begin{document}

\title{Optical and X-ray Properties of the Swift BAT-detected AGN}


\author{Lisa M. Winter}{
  address={Hubble Fellow}
  ,altaddress={University of Colorado, Boulder, CO, USA}
}

\author{Richard Mushotzky}{
  address={University of Maryland, College Park, MD, USA}
}

\author{Karen Lewis}{
 address={Dickinson College, Carlisle, PA, USA}
}

\author{Sylvain Veilleux}{
  address={University of Maryland, College Park, MD, USA}
}

\author{Michael Koss}{
  address={University of Maryland, College Park, MD, USA}
}

\author{Brian Keeney}{
  address={University of Colorado, Boulder, CO, USA}
}

\begin{abstract}
 The Swift Gamma-Ray Burst satellite has detected a largely unbiased towards absorption sample of local ($<z> \approx 0.03$) AGN, based solely on their 14--195 keV flux.  In the first 9 months of the survey, 153 AGN sources were detected.  The X-ray properties in the 0.3--10 keV band have been compiled and presented based on analyses with XMM-Newton, Chandra, Suzaku, and the Swift XRT (Winter et al. 2009).  Additionally, we have compiled a sub-sample of sources with medium resolution optical ground-based spectra from the SDSS or our own observations at KPNO.  In this sample of 60 sources, we have classified the sources using standard emission line diagnostic plots, obtained masses for the broad line sources through measurement of the broad H$\beta$ emission line, and measured the [OIII] 5007\AA~luminosity of this sample.  Based on continuum fits to the intrinsic absorption features, we have obtained clues about the stellar populations of the host galaxies.  We now present the highlights of our X-ray and optical studies of this unique sample of local AGNs, including a comparison of the 2--10 keV and 14--195 keV X-ray luminosities with the [OIII] 5007\AA~luminosity  and the implications of our results towards measurements of bolometric luminosities.
\end{abstract}

\maketitle


\section{Introduction}

Searching the sky in the very hard X-rays (14--195\,keV), the Swift Burst Alert Telescope (BAT)
provides an unbiased view of local Compton thin active galactic nuclei sources (AGNs).  From the first 9 months of the BAT survey,
a sample of 153 AGNs were identified with BAT fluxes of $\ge 2 \times 10^{-11}$\,ergs\,s$^{-1}$\,cm$^{-2}$  \cite{2008ApJ...681..113T}.  These sources are nearby, with average redshifts of 0.03, and bright, offering an excellent sample for multi-wavelength follow-ups.

Following upon our analysis of the X-ray properties in the 0.3--10\,keV band \cite{2009ApJ...690.1322W}, we have obtained and analyzed the optical spectra of 60 of the 9-month sources
\cite{Winter-optical}.  These spectra include 27 from archived Sloan Digital Sky Survey (SDSS) observations and 40 from our follow-up observations with the Kitt Peak National Observatory's 2.1-m telescope.  The sources include 80\% of the `northern' BAT-detected AGN, comprising a representative sub-sample of the 9-month Swift AGN catalog.  This sample consists of 50\% broad line/X-ray unabsorbed and 50\% narrow line/X-ray absorbed sources.

The emission line and host galaxy absorption/continuum properties were measured for each of the optical spectra.  Among our results we found that the emission line diagnostics indicate that the majority of the BAT-detected sources are optically identified as Seyferts.  We also find that the host galaxy properties, particularly from the narrow line sources where the regions including the Ca\,II break and H$\delta$ absorption are unaffected by broad Balmer emission lines, are consistent with late type galaxies -- the same result found from narrow line sources, within the same redshift range as the Swift BAT sources, in the SDSS studies \cite{2003MNRAS.346.1055K}.  With such a comprehensive database of the X-ray and optical properties of the Swift BAT AGN sources, we can now begin to compare the properties in both bands. 
  
\begin{figure}
\includegraphics[height=.3\textheight]{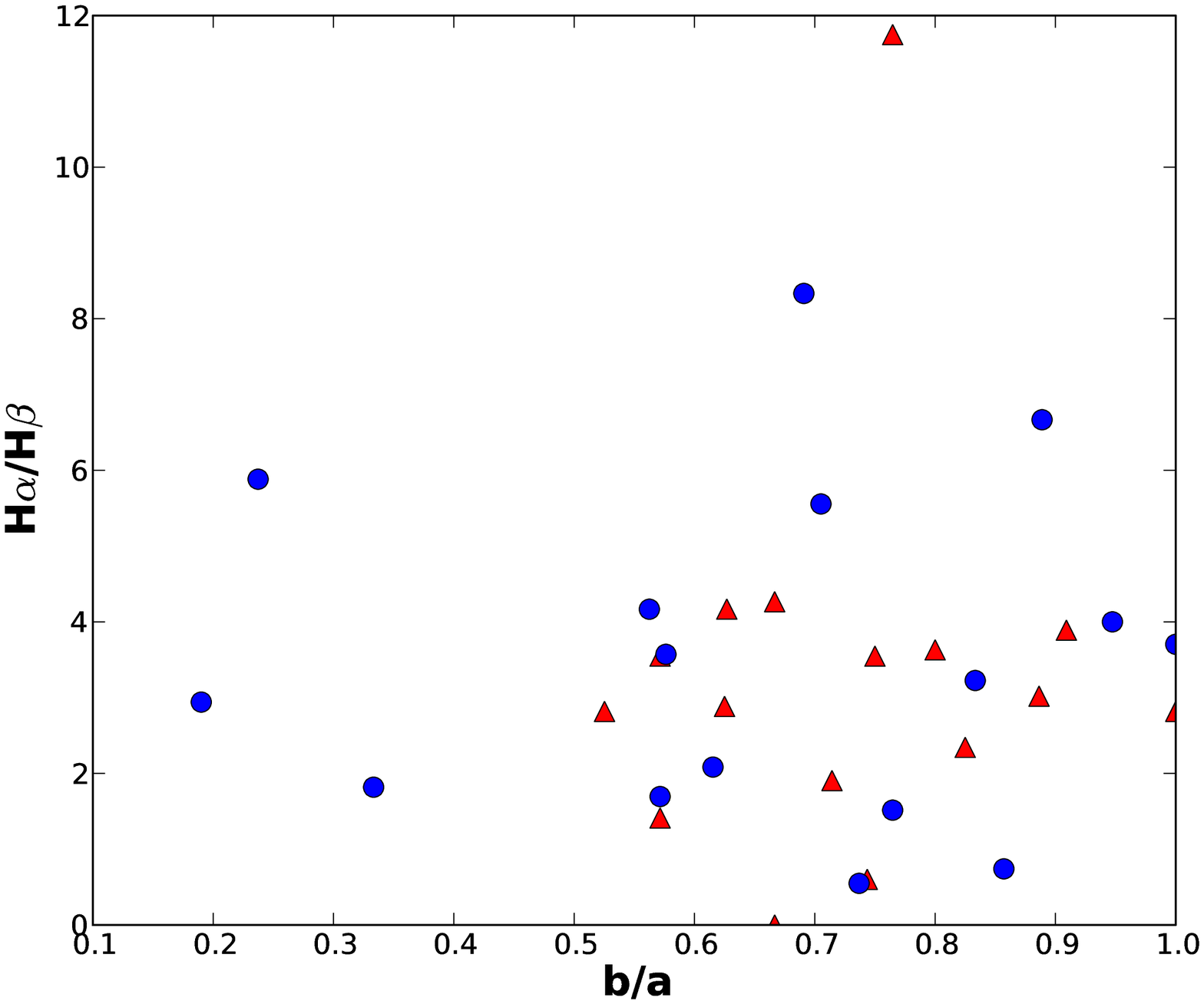}
\includegraphics[height=.3\textheight]{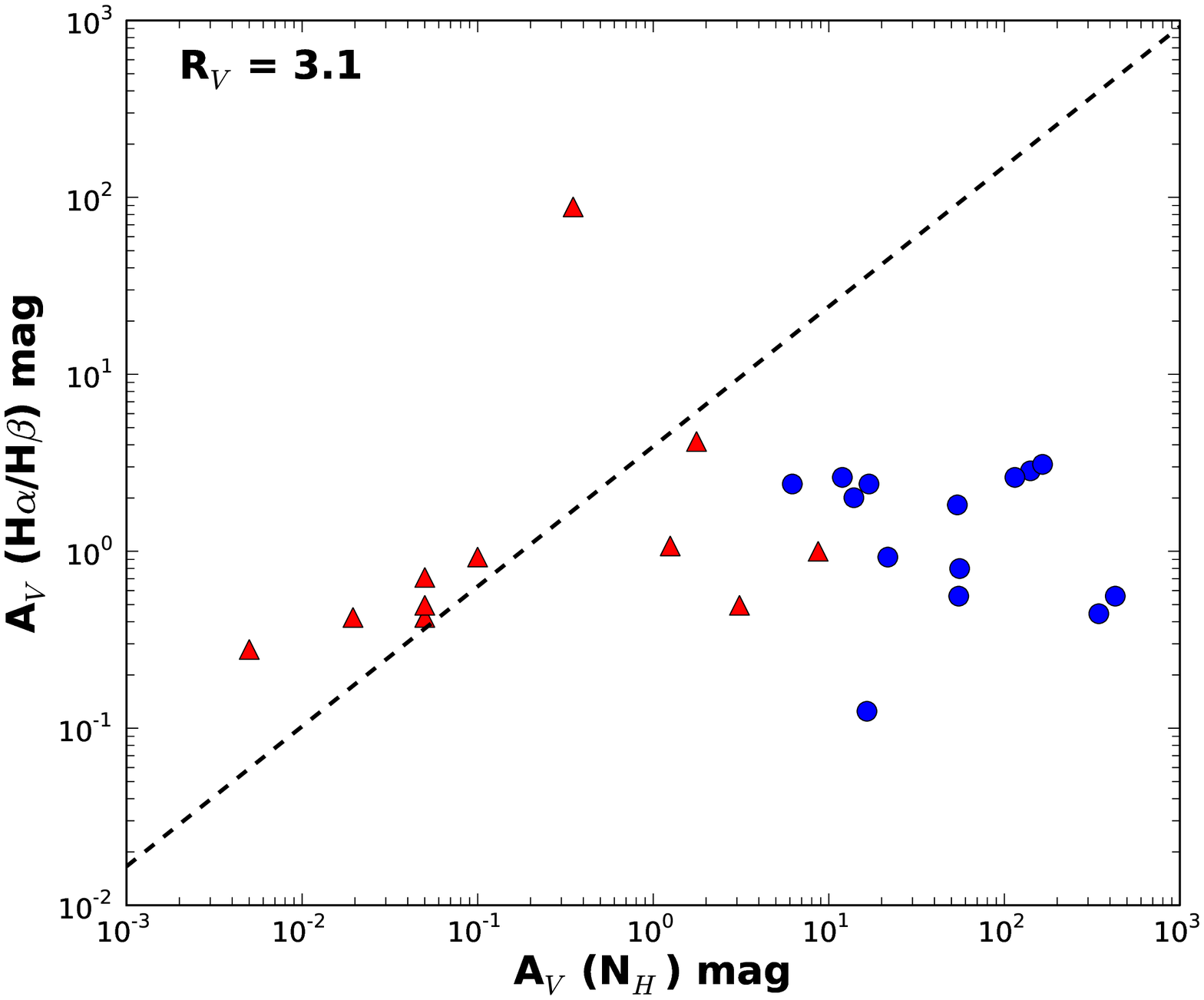}\\

  \caption{{\it Left:} Optical reddening, measured through the narrow Balmer line decrement, versus host galaxy inclination (minor axis/major axis) is shown.  There is no correlation for either the broad line (triangles) or narrow line (circles) sources.  This shows that the optical reddening originates closer to the central black hole. {\it Right:} Optical extinction is shown as a function of the X-ray extinction, derived from the X-ray hydrogen column density ($N_{\rm H} = 2 \times 10^{21} A_{V}$). No correlation is seen for the narrow line sources.  However, these values are linearly correlated for the broad line sources.}
\end{figure}

\section{Optical and X-ray Extinction}
The narrow line Balmer decrement, H$\alpha$/H$\beta$, is an indicator of the optical reddening affecting a source.  From our analysis of the optical spectra, we found that the amount of reddening is lower for the broad line sources than the narrow line sources.  We find that the optical reddening, like the X-ray measured hydrogen column density, is not a function of the host galaxy inclination angle (Figure 1a).  This shows that for both the optical and X-ray extinction, the primary source of absorbing gas and dust is not the host galaxy but from a region closer to the central black hole.

We also find that the reddening in the optical is correlated with reddening estimated from the X-ray derived column densities, but only for the Seyfert 1s (Figure 1b).   For the more heavily absorbed X-ray sources, the estimated X-ray extinction is greater than the optical extinction by a factor of 10s--100s.  Such differences in the optical and X-ray extinction of narrow line sources have been noted in other samples and could be caused by dust dominated with large grains, which absorb without causing much spectral reddening \cite{2001AA...365...28M}, an X-ray absorber that does not cover the optical emission region, or the presence of ionized absorbers, which absorb X-rays but have no opacity in other bands.


\section{X-ray and [O III] Luminosities}
The luminosity in the prominent [O III] $\lambda 5007$\AA~emission line has often been used as a proxy for the AGN bolometric luminosity \cite{2005ApJ...634..161H}.  However, ionization from hot young stars as well as galactic extinction can affect the measurements of [O III].  Since the BAT detects AGN in the 14--195\,keV band, the energies probed are not affected by stellar light and obscuration from all but the highest column densities (i.e. $\log N_{\rm H} > 24$).  When we compare the BAT luminosities of the sources in our sample with our extinction and stellar continuum corrected [O III] luminosities, we find only a weak correlation.  Further, there is a lot of scatter in the distribution, particularly for the absorbed sources.  This is also true in the observed [O III] luminosities.  Therefore, our results confirm those previously drawn from a smaller sample of Swift AGN \cite{2008ApJ...682...94M}, showing that caution needs to be applied when using [O III] as an indicator of bolometric luminosity.

In addition to X-ray measurements in the BAT band, we were also able to compare the [O III] luminosities with those obtained from spectral fits in the soft (0.5--2\,keV) and hard (2--10\,keV) X-ray bands.  We found that the Seyfert 1s had similar weak correlations between [O III] and all of the measured X-ray bands (BAT, soft, and hard).  For the Seyfert 2s, however, the best correlation was seen between [O III] and the soft X-ray band (Figure 2a).  A similar result was also seen in AGN selected from the XMM slew survey \cite{2009ApJ...705..454N}.  As shown in Figure 2b, the ratio of the soft to hard flux can also be interpreted as the scattering fraction, as presented in the talk by Terashima.  Sources with low scattering fractions have small opening angles in the torus.  Such sources are heavily absorbed and tend to be missed in both optical and soft X-ray surveys.

\begin{figure}
\includegraphics[height=.3\textheight]{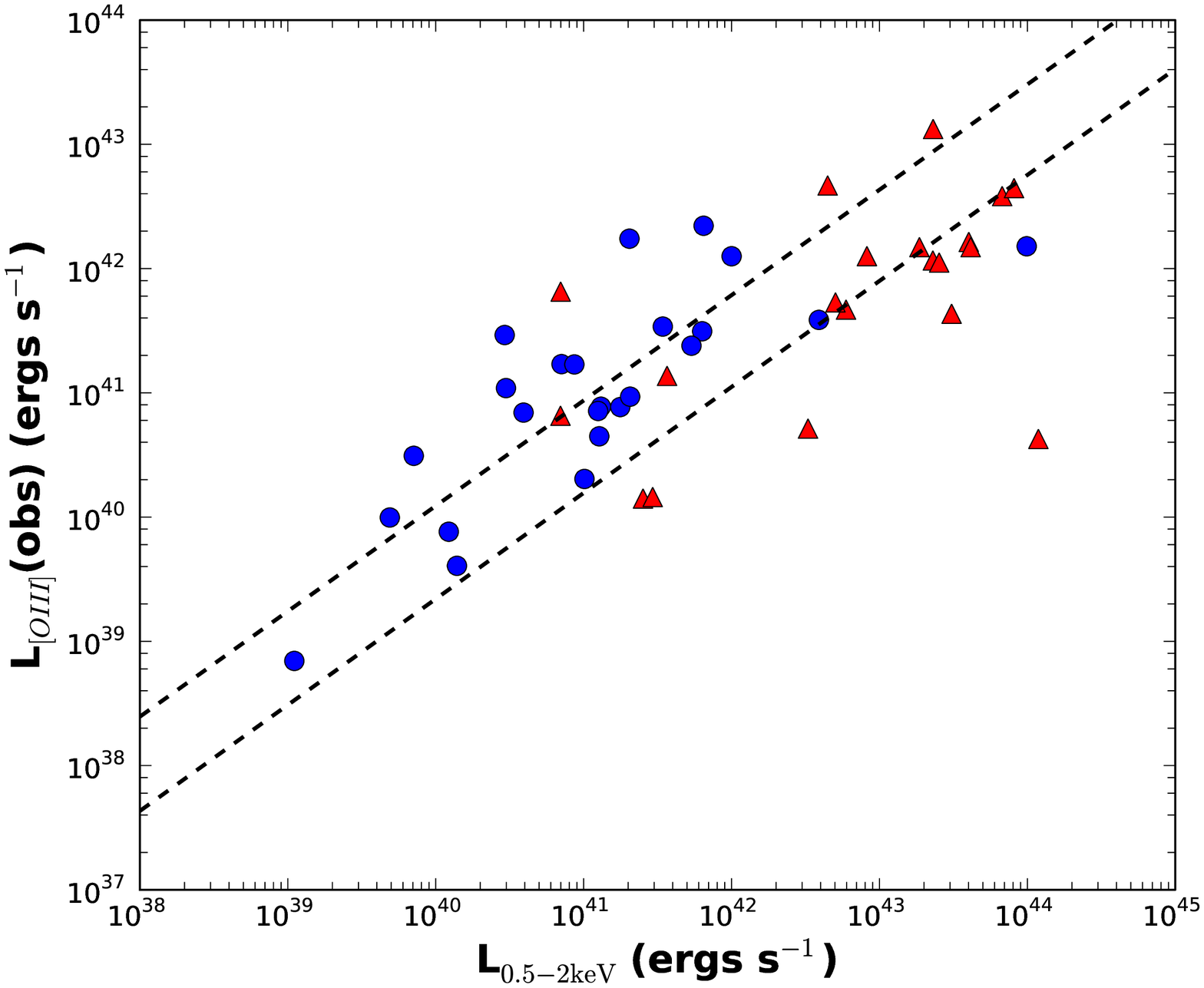}
\includegraphics[height=.3\textheight]{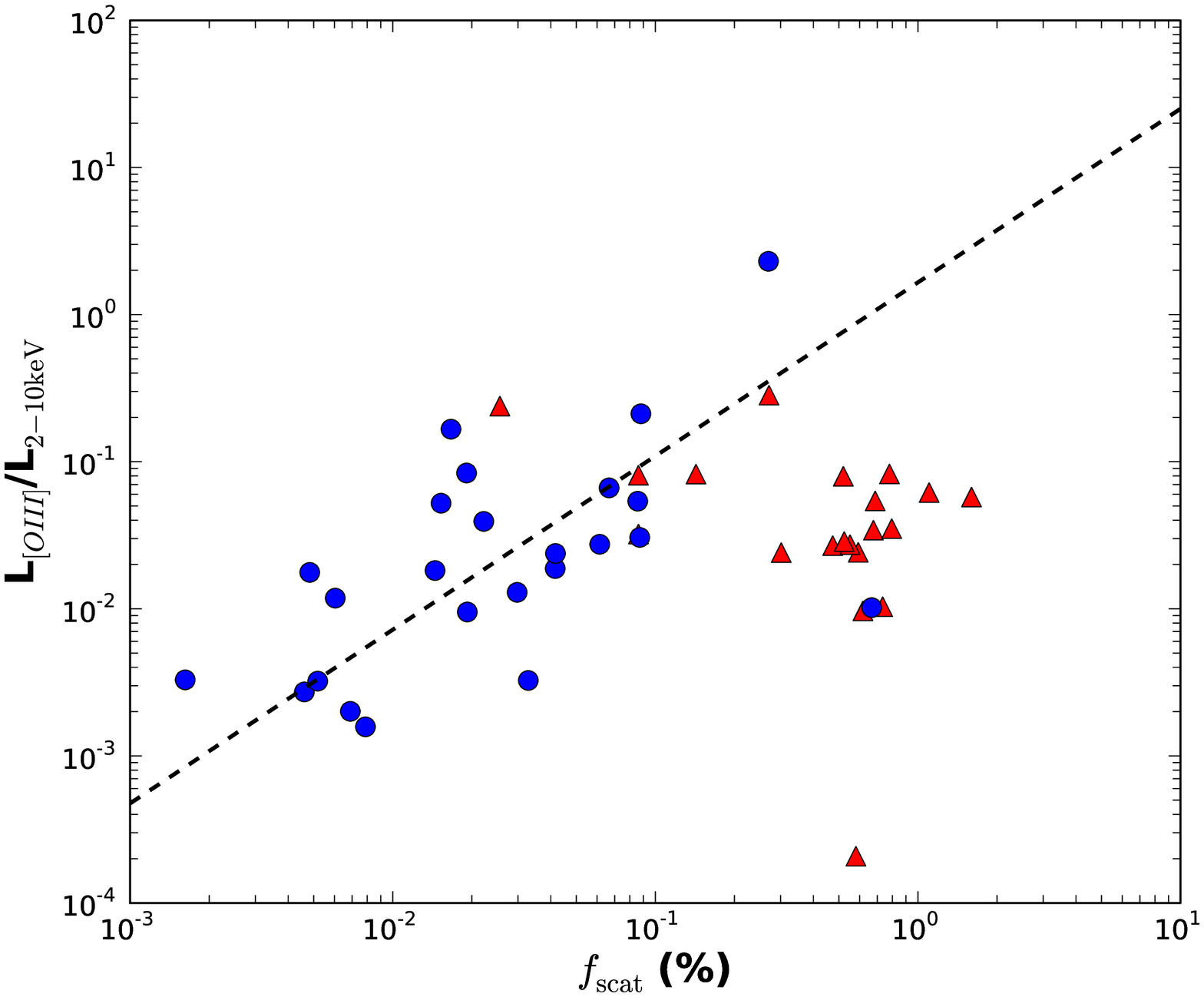}\\

  \caption{{\it Left:} For the narrow line sources (circles), we find that the greatest correlation exists between the [O III] $\lambda 5007$\AA~luminosity and the observed 0.5--2\,keV luminosity. {\it Right:} This correlation is reminiscent of the findings presented by Terashima, where the scattering fraction of the torus is small in these obscured sources, which have been missed in optically selected samples.}
\end{figure}

\section{Multi-wavelength Studies of the BAT Sample}
As we have shown, the persistent brightness and low redshift of the Swift BAT detected AGN make them an excellent sample for multi-wavelength follow-ups.  We have completed an analysis of the 0.3--10\,keV X-ray spectral properties for the 153 sources in the 9-month catalog \cite{2008ApJ...681..113T}.  We have also completed an analysis of the optical spectra ($\approx 3700$--$7500$\AA) of 60 of these sources.  With continued study of these sources, we hope to gain a full understanding of the properties of local AGNs and their host galaxies.

Towards this end, several other studies of the BAT AGNs are already underway.  Among these, 
a comparison of the IR [O\,IV], optical [O\,III], and X-ray 2--10\,keV luminosity are presented in \citet{2008ApJ...682...94M} for a sample of 40 BAT AGNs.  Simultaneous optical-to-X-ray spectral energy distributions are analyzed for 26 of the BAT AGNs in \citet{2009arXiv0907.2272V}.  Additionally, some details of the optical host properties are presented in \citet{2009ApJ...690.1322W} as well as 
\citet{2009ApJ...692L..19S}.  Further, the results of a full analysis of the optical colors and morphologies are being compiled in Koss {\it et al.} (in prep) and the Spitzer-based IR properties will be presented in Weaver {\it et al.} (in prep).

\begin{theacknowledgments}
  L. M. W. is supported by NASA through Hubble Fellowship grant \# HF--51263.01 awarded by the Space Telescope Science Institute, which is operated by the Association of Universities for Research in Astronomy, Inc., for NASA, under contract NAS 5-26555.
\end{theacknowledgments}



\bibliographystyle{aipproc}   

\bibliography{../../MyBibtex.bib}

\begin{thebibliography}{10}
\expandafter\ifx\csname natexlab\endcsname\relax\def\natexlab#1{#1}\fi
\providecommand{\enquote}[1]{``#1''}
\expandafter\ifx\csname url\endcsname\relax
  \def\url#1{\texttt{#1}}\fi
\expandafter\ifx\csname urlprefix\endcsname\relax\def\urlprefix{URL }\fi
\providecommand{\eprint}[2][]{\url{#2}}

\bibitem[{Tueller} et~al.(2008)]{2008ApJ...681..113T}
J.~{Tueller}, R.~F. {Mushotzky}, S.~{Barthelmy}, J.~K. {Cannizzo},
  N.~{Gehrels}, C.~B. {Markwardt}, G.~K. {Skinner}, and L.~M. {Winter},
  \emph{\apj} \textbf{681}, 113--127 (2008), \eprint{arXiv:0711.4130}.

\bibitem[{Winter} et~al.(2009{\natexlab{a}})]{2009ApJ...690.1322W}
L.~M. {Winter}, R.~F. {Mushotzky}, C.~S. {Reynolds}, and J.~{Tueller},
  \emph{\apj} \textbf{690}, 1322--1349 (2009{\natexlab{a}}),
  \eprint{0808.0461}.

\bibitem[{Winter} et~al.(2009{\natexlab{b}})]{Winter-optical}
L.~M. {Winter}, K.~T. {Lewis}, M.~{Koss}, S.~{Veilleux}, B.~{Keeney}, and
  R.~{Mushotzky}, \emph{\apj} \textbf{(submitted)} (2009{\natexlab{b}}).

\bibitem[{Kauffmann} et~al.(2003)]{2003MNRAS.346.1055K}
G.~{Kauffmann}, T.~M. {Heckman}, C.~{Tremonti}, J.~{Brinchmann}, S.~{Charlot},
  S.~D.~M. {White}, S.~E. {Ridgway}, J.~{Brinkmann}, M.~{Fukugita}, P.~B.
  {Hall}, {\v Z}.~{Ivezi{\'c}}, G.~T. {Richards}, and D.~P. {Schneider},
  \emph{\mnras} \textbf{346}, 1055--1077 (2003),
  \eprint{arXiv:astro-ph/0304239}.

\bibitem[{Maiolino} et~al.(2001)]{2001AA...365...28M}
R.~{Maiolino}, A.~{Marconi}, M.~{Salvati}, G.~{Risaliti}, P.~{Severgnini},
  E.~{Oliva}, F.~{La Franca}, and L.~{Vanzi}, \emph{\aap} \textbf{365}, 28--36
  (2001), \eprint{arXiv:astro-ph/0010009}.

\bibitem[{Heckman} et~al.(2005)]{2005ApJ...634..161H}
T.~M. {Heckman}, A.~{Ptak}, A.~{Hornschemeier}, and G.~{Kauffmann}, \emph{\apj}
  \textbf{634}, 161--168 (2005), \eprint{arXiv:astro-ph/0507674}.

\bibitem[{Mel{\'e}ndez} et~al.(2008)]{2008ApJ...682...94M}
M.~{Mel{\'e}ndez}, S.~B. {Kraemer}, B.~K. {Armentrout}, R.~P. {Deo}, D.~M.
  {Crenshaw}, H.~R. {Schmitt}, R.~F. {Mushotzky}, J.~{Tueller}, C.~B.
  {Markwardt}, and L.~{Winter}, \emph{\apj} \textbf{682}, 94--103 (2008),
  \eprint{0804.1147}.

\bibitem[{Noguchi} et~al.(2009)]{2009ApJ...705..454N}
K.~{Noguchi}, Y.~{Terashima}, and H.~{Awaki}, \emph{\apj} \textbf{705},
  454--467 (2009), \eprint{0910.0965}.

\bibitem[{Vasudevan} et~al.(2009)]{2009arXiv0907.2272V}
R.~V. {Vasudevan}, R.~F. {Mushotzky}, L.~M. {Winter}, and A.~C. {Fabian},
  \emph{ArXiv e-prints}  (2009), \eprint{0907.2272}.

\bibitem[{Schawinski} et~al.(2009)]{2009ApJ...692L..19S}
K.~{Schawinski}, S.~{Virani}, B.~{Simmons}, C.~M. {Urry}, E.~{Treister},
  S.~{Kaviraj}, and B.~{Kushkuley}, \emph{\apjl} \textbf{692}, L19--L23 (2009),
  \eprint{0901.1663}.

\end{thebibliography}

\IfFileExists{\jobname.bbl}{}
 {\typeout{}
  \typeout{******************************************}
  \typeout{** Please run "bibtex \jobname" to optain}
  \typeout{** the bibliography and then re-run LaTeX}
  \typeout{** twice to fix the references!}
  \typeout{******************************************}
  \typeout{}
 }

\end{document}